# An Approach to *Ad hoc* Cloud Computing

Graham Kirby, Alan Dearle, Angus Macdonald
School of Computer Science
University of St Andrews, St Andrews,
Fife, Scotland KY16 9SX
+44 1334 463253
{graham,al,angus}@cs.st-andrews.ac.uk

Alvaro Fernandes
School of Computer Science
University of Manchester, Oxford Road,
Manchester, United Kingdom M13 9PL
+44 161 306 9280
a.fernandes@manchester.ac.uk

## ABSTRACT
We consider how underused computing resources within an enterprise may be harnessed to improve utilization and create an elastic computing infrastructure. Most current cloud provision involves a data center model, in which clusters of machines are dedicated to running cloud infrastructure software. We propose an additional model, the *ad hoc* cloud, in which infrastructure software is distributed over resources harvested from machines already in existence within an enterprise. In contrast to the data center cloud model, resource levels are not established a priori, nor are resources dedicated exclusively to the cloud while in use. A participating machine is not dedicated to the cloud, but has some other primary purpose such as running interactive processes for a particular user. We outline the major implementation challenges and one approach to tackling them.

## 1. INTRODUCTION
Computational and storage resources within organizations are often under-utilized. This is likely to increase with further adoption of cloud services. A volunteer cloud infrastructure, supporting what we term *ad hoc* cloud computing, would allow cloud services to run on existing heterogeneous hardware.

If available, such infrastructure could improve organizations' resource utilization while offering some of the benefits of more conventional public and private clouds. This could yield significant cost savings. The model is analogous to volunteer computing as exemplified by Condor [21] and BOINC [22], although it poses considerable additional implementation challenges.

In particular, we are interested in increasing utilization of general-purpose computers in offices and laboratories. As a motivating example, the (small) University of St Andrews operates in the region of ten thousand machines in offices and labs. In aggregate, their unused processing and storage capacity represent a major untapped computing resource.

The recent Draft NIST Working Definition of Cloud Computing [20] defines both public and private cloud models. Both may be termed *data center models*, in which clusters of machines are dedicated to running cloud infrastructure software. We propose to introduce an additional deployment model, the *ad hoc* cloud, in which infrastructure software is distributed over resources harvested from machines already in use. By *ad hoc* we mean that the set of machines comprising the cloud changes dynamically, as does the proportion of each machine's computational and storage resources that can be harnessed at a given point in time. Thus, in contrast to the data center cloud model, resource provisioning levels are not established a priori, nor are resources committed exclusively to the cloud while in use. A participating machine is not dedicated to the cloud, but has some other primary purpose such as running interactive processes for a particular user, albeit often for a small proportion of the time. One of the most important research issues is how to reduce the impact of cloud operations on such processes to an acceptable level.

The availability of *ad hoc* clouds could yield various benefits to individual enterprises. Firstly, it could reduce the numbers of machines that need to be purchased. Such costs are borne directly by enterprises employing private clouds, and indirectly by those using external cloud providers[1].

The use of *ad hoc* clouds could also reduce the need for specialized infrastructure for resilience, such as redundant power and cooling systems, battery backup, etc. This represents 25% of data center costs [13]. Rather than ensuring resilience of a small number of physical buildings, the grain of resilience could be expanded by using more widely distributed machines and tolerating individual building failures.

*Ad hoc* clouds could reduce overall power consumption. One factor is a reduction in the total number of machines required—significant since the energy cost of manufacture for a computer has been estimated as four times that used during its lifetime [23]. Another is that since machines comprising an *ad hoc* cloud infrastructure are situated in working spaces, the power consumed is partially offset (in temperate climates) by a reduction in the power required for heating. Conversely, machines are housed at lower densities than in data centers, so less active cooling is required.

A similar idea, that of Nebulas, was proposed in [4]. Here we outline a specific approach to developing such *ad hoc* infrastructure. Section 2 outlines requirements and the principal implementation challenges; Section 3 surveys related work; Section 4 describes our proposed approach to the problem.

## 2. RELATED WORK
The approach we describe can be compared and contrasted with grid and volunteer computing, and provider-specific clouds.

Grid computing emerged principally to address requirements from e-Science, in which there was a growing need for software platforms that supported sharing of resources to support collaborative data analysis in computationally intensive science. Grid computing provides facilities for the sharing of computational resources, often across administrative domains, with a view to enabling effective collaboration between the owners of data or computational resources. To support such capabilities, grid toolkits (e.g. Globus [9]) provide core facilities that support operating system style functionalities such as file access, job execution and authentica-tion, across heterogeneous platforms. These can be used to support higher-level services such as distributed file systems (e.g. SRB [18]), workflow execution (e.g. Condor-G [12]) and workflow management (e.g. Pegasus [8]). Higher-level grid functionalities, such as abstract workflow specification in Pegasus, often make use of lower-level platforms (e.g. Pegasus uses Condor-G for managing dependencies between multiple jobs, which in turn uses Globus for job execution and file replica management).

[1] It has been estimated that 45% of data center costs are incurred in purchasing servers [13].

Grids have been a focus of considerable research, development and commercial activity for a decade, giving rise to a range of approaches and emphases. A significant portion of the work focuses on connecting high-end, heterogeneous computational resources across multiple administrative domains, with a view to supporting virtual organisations, for example [11]. This emphasis has not been substantially changed by the move towards service-oriented grid architectures (e.g. [10]), in which resources are virtualised as web services, and thus grid functionalities are made available as part of a wider, service-oriented architecture. As such, the grid community has considerable experience in the development of techniques for providing abstractions over heterogeneous platforms.

The cloud vision has elements in common with the objectives of grid computing, in particular a reduction in costs through resource sharing, and improvements in flexibility and reliability. However, different starting points have given rise to differing architectures and emphases. Broadly, grids have sought to support coordinated use of distributed resources for carrying out computationally intensive tasks for modest numbers of users, whereas clouds have focused on coordinated use of largely centralised resources for large numbers of less demanding requests from distributed users.

Volunteer computing (VC), sometimes described as a desktop grid, uses individual users' machines to perform computationally intensive tasks. It is particularly suited for 'embarrassingly parallel' problems, e.g. SETI@home, one of a number of popular projects based on the BOINC framework [22]. *Ad hoc* clouds share the goal of 'stealing cycles' from user machines, but target more diverse applications. They can be viewed as offering the resource utilisation benefits of VC while avoiding the limitations of low or fluctuating volunteering rates, and providing the elasticity to workloads that make the cloud vision appealing.

The Condor platform [21] also supports resource harvesting for highly parallel tasks. However, Condor is concerned with task scheduling whereas our approach targets a more general application-hosting model, in particular the support of interactive and data-centric applications.

The best-known examples of Cloud computing, such as those of Amazon, Google, Yahoo! and Microsoft, have several aspects in common. For example, early clouds have been developed to support scale-out: the execution of large numbers of typically constrained requests over potentially huge data sets. This in turn has led to the development of simplified but scalable computational models, such as Google's MapReduce framework [6], which provides a simple model for distributing highly parallelisable problems over large machine clusters. The implementation abstracts over the details of distributing input data to individual machines and collecting results, and has been widely adopted by other cloud platforms, which often make use of Hadoop [3], an open-source implementation of the MapReduce model. MapReduce, in common with early cloud data management platforms such as Amazon's Simple Storage Service (S3) and SimpleDB [2], and Google's Bigtable storage system [5], provides carefully constrained capabilities. Google AppEngine also provides a constrained model, specifically targeting web applications.

At a lower level of abstraction, Amazon Elastic Compute Cloud (EC2) [2] allows an application to be structured as a set of potentially communicating virtual machine instances. The term 'elastic' refers to the flexibility with which instances may be created and discarded dynamically, allowing the computing resources allocated to applications to scale as required.

Early support for cloud service developers, then, offers two distinct styles: high-level APIs that significantly constrain service structure, and low-level machine virtualisation that gives almost complete freedom but provides little assistance with partitioning and managing the service across virtualised instances.

From the perspective of the service provider, constrained service provision offers distinct benefits, as discussed for cloud data services in the Claremont Report on Database Research:

*"Early cloud data services offer an API that is much more restricted than that of traditional database systems, with a minimalist query language and limited consistency guarantees. This pushes more programming burden on developers, but allows cloud providers to build more predictable services, and to offer service level agreements that would be hard to provide for a full-function SQL data service. More work and experience will be needed on several fronts to explore the continuum between today's early cloud data services and more full-functioned but probably less predictable alternatives."* [1]

More recently, Amazon and Microsoft have introduced full relational database facilities, while the Windows Azure platform [17] offers a rather richer set of APIs to programmers. The cloud infrastructure remains targeted at dedicated servers.

# 3. RESEARCH ISSUES

*Ad hoc* clouds can be thought of as a generalization of public or private data center clouds, in which certain assumptions are relaxed. These include the degree of homogeneity and availability of servers, and the presence of non-cloud processes on cloud hosts. *Ad hoc* clouds could host and coordinate services that are more diverse than those currently associated with high-level cloud APIs, while providing the service developer with richer support for service partitioning and management than machine virtualisation approaches. They would operate over shared, heterogeneous resources, thus giving rise to requirements for more complex automatic management and more sophisticated quality of service handling.

We identify a resulting set of issues that would need to be addressed:

**Core functionality:**

- What are the architectural requirements for an *ad hoc* cloud infrastructure?
- What mechanisms are needed to allow convenient access to services, without single points of failure?

- How can membership of the set of machines in an *ad hoc* cloud be controlled? In which situations should an *ad hoc* cloud be scaled out or contracted?
- For some application classes, current cloud approaches scale well in stable environments—to what extent can these restrictions be relaxed while retaining scalability?

**Automatic adaptation:**

- Can speculative plans for actions that might improve *ad hoc* cloud operation be generated automatically?
- What techniques are needed in order to model *ad hoc* cloud behavior to enable useful estimates of the consequences of possible autonomic reconfiguration?
- How should planning and modeling processes be coordinated? Where are they executed, and how are their resources allocated?
- To what extent can planning decisions be improved using measurements and predictions of previous, current and future workloads?
- What model calibration techniques are needed? In what situations do phase changes in user behavior or the environment cause a previously accurate model to diverge from reality, and how can this be handled?
- How can the characteristics of a particular application be taken into account in determining how the ad hoc cloud adapts to support it?

**Quality of service:**

- To what extent can useful QoS guarantees be delivered to *ad hoc* cloud clients while limiting disruption for machine owners to acceptable levels?
- In what way does the class of computation supported by a cloud influence the quality of service guarantees that can be provided?
- What are the appropriate forms for expression of high-level policy goals, for an entire ad hoc cloud, and for specific services?
- Can high-level goals be translated automatically to corresponding concrete actions?
- How should measured low-level properties be aggregated for reporting in terms of high-level goals?
- Under what circumstances can conflicting policies be detected and automatically resolved?
- What mechanisms can be used to coordinate potentially complementary services (e.g. block storage, file systems, databases), so that they align with one another rather than competing unnecessarily for resources?

# 4. OUR APPROACH

An *ad hoc* cloud should be self-managing in terms of resilience, performance and balancing potentially conflicting policy goals. For resilience it should maintain service availability in the presence of membership churn and failure. For performance it should be self-optimizing, taking account of quality of service requirements. It should be acceptable to machine owners, by minimising intrusiveness and supporting appropriate security and trust mechanisms.

We identify several desirable features for the general *ad hoc* cloud architecture:

- Agnostic as to service type: the approach can be applied to different styles of cloud service, for example infrastructure, platform or application as service [20].
- Agnostic as to means of control: the approach allows different forms of autonomic decision making to be deployed at different points in the architecture.
- Agnostic as to grain of control: autonomic behaviour may be coarse or fine-grained at different points. For example, a dispatcher may balance load only as requests leave queues, or may change resource allocations for running jobs.

## 4.1 Core Cloud Functionality

There exist successful architectures for large-scale cloud computing in the data center style [2, 3, 5, 6]. The principal additional challenges in supporting *ad hoc* clouds lie in accommodating highly dynamic machine membership, and allowing cloud computations to co-exist satisfactorily with non-cloud processes. Although data center clouds deal with machine failures automatically, the churn will be significantly higher in *ad hoc* clouds, arising from more frequent rebooting of personal machines and the frequent unavailability of portable devices. Machines may also become unavailable to the *ad hoc* cloud for unpredictable periods, even though they remain connected and functioning, due to the higher priority of fluctuating user workloads. Data center clouds do not support co-existence of cloud and user processes; an *ad hoc* cloud architecture must support monitoring of impact on user processes, rapid relocation or shut-down of cloud processes, and modelling of cloud computation to allow sensible initial placement of cloud processes.

Here we sketch a possible architecture as a starting point. We define an *ad hoc* cloud as the union of a set of *cloudlets*, each of which provides a particular service or application. A cloudlet service may be specified and accessed via Web Services, or any other convenient protocol. Each cloudlet runs on a potentially dynamically changing set of physical machines. A given machine may host parts of multiple cloudlets.

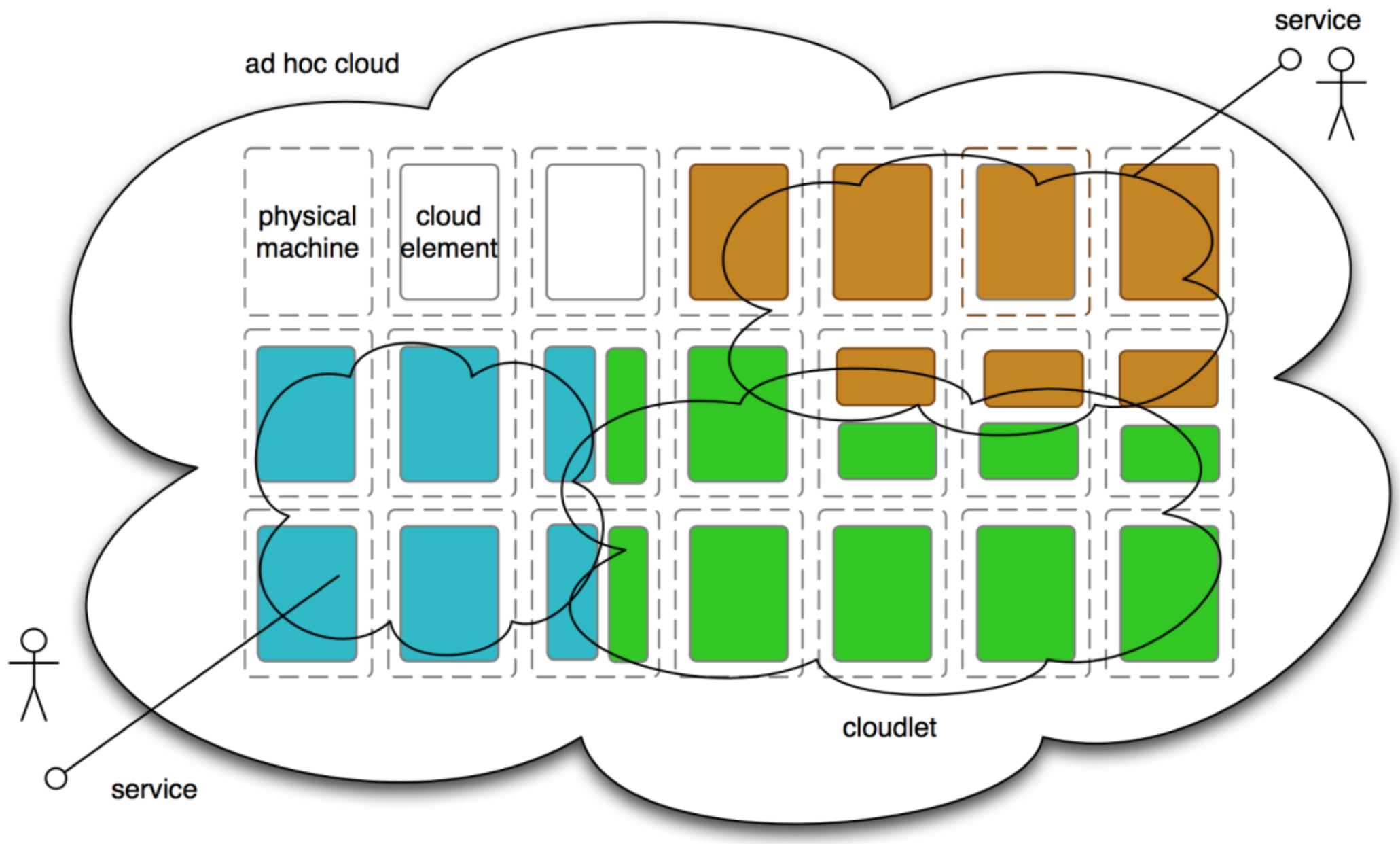

**Figure 1. Cloudlets and Cloud Elements**

The software running on a particular machine, contributing to a particular cloudlet, is termed a *cloud element*. A cloudlet may be expanded or contracted by altering the number of machines, and hence cloud elements, assigned to it. The cloud elements comprising a given cloudlet communicate with one another to coordinate their activity. The cloud elements within a cloudlet may be, but need not be, homogeneous in terms of their functionality. This structure is illustrated in Figure 1 (although it may suggest that the cloud elements assigned to each particular cloudlet run on physically close machines, this is for convenience of drawing only, and no such restriction is imposed by the architecture).

Each physical machine available to the *ad hoc* cloud may host a number of cloud elements, each assigned to a different cloudlet. The machine runs cloud infrastructure software, which supports secure creation, management and destruction of cloud elements. Finally, the machine also executes non-cloud processes for the primary user. This is shown in Figure 2.

The cloud infrastructure contains an *element manager* for creating and destroying cloud elements. It also contains a *modeller/manager*—which interacts with the host operating system in order to monitor effects of the local cloud elements on the user processes, and vice versa—and a *broker* and *dispatcher*, whose functions are described in the QoS section.

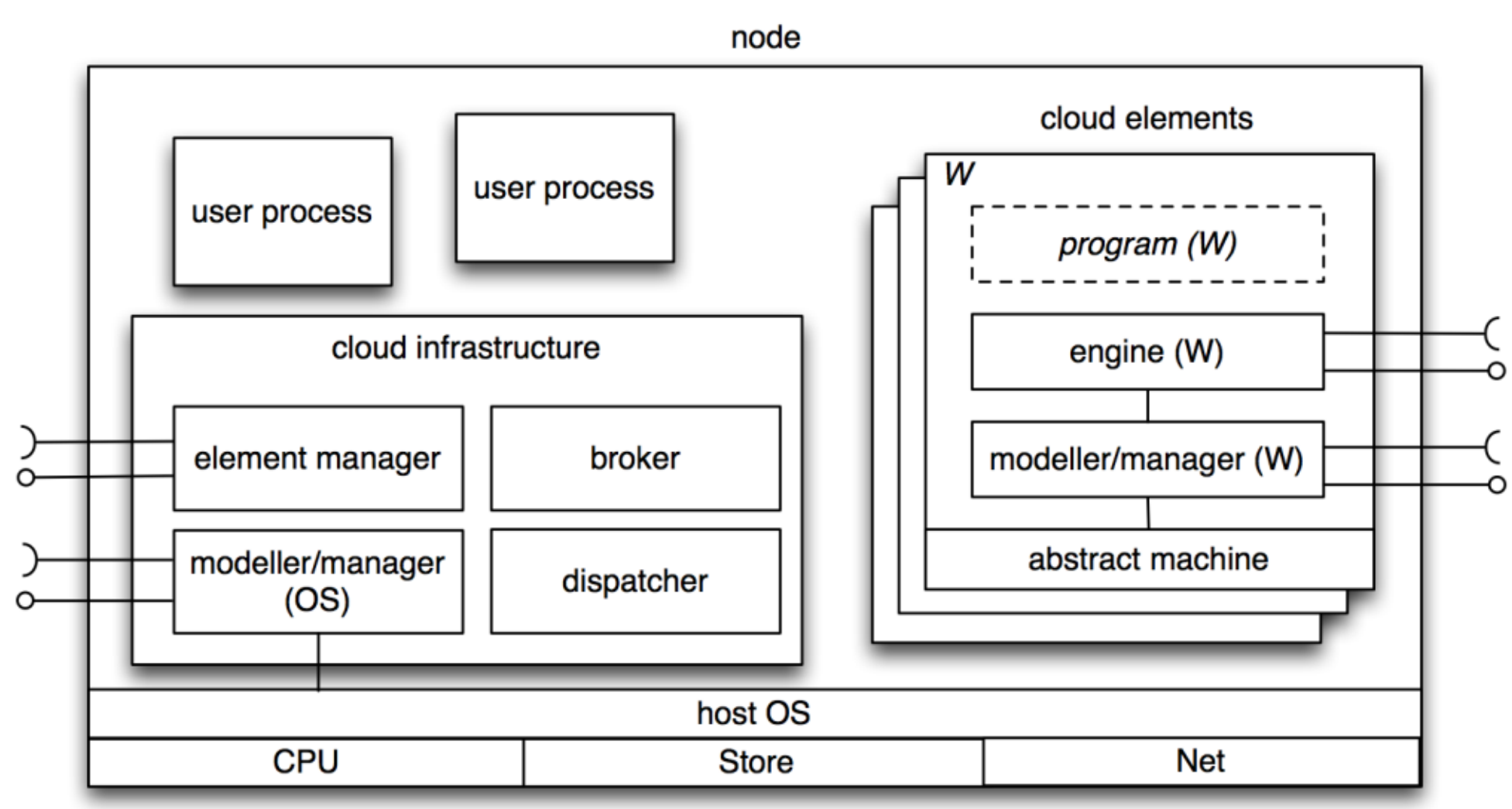

**Figure 2. Node Structure**

Each cloud element running on a machine contains an *engine* capable of running the class of computations appropriate to its cloudlet. The engine in the diagram is labelled *W* to signify that it can execute a particular class of workloads *W*, corresponding to the cloudlet functionality. An engine may provide application functionality directly via a user-level API, or support a further layer of application software loaded onto it. For example, one engine might provide a SQL API that accepts user queries directly; other engines might provide *MapReduce* functionality, or Java or JavaScript interpreters. In all but the first case a corresponding program would also be loaded. Each engine runs on an abstract machine. This might be a VMware-style virtual machine, a Java-style VM, or something else tailored to the target computation class. Whatever it is, it must provide sufficient isolation of the element from the non-cloud processes on the node.

The cloud element also contains its own modeller/manager, which has knowledge of the semantics of *W*, and a cost model that allows reasoning about how the computation will be executed. The purpose of the modeller/manager is to control the operation of its associated engine such as to minimise disruption to user processes, to optimise its contribution to cloudlet functionality within such constraints, and to publish information to support effective deployment and adaptation of cloud elements.

The infrastructure modeller/manager provides a conduit for communication between cloud element modeller/managers and the host OS. For example, it allows cloud element modeller/managers to be aware of the current resource demands of non-cloud user processes, and hence endeavour to avoid undue disruption. Communication between cloud element modeller/managers is necessary in order to coordinate operations across a cloudlet.

## 4.2 Automatic Adaptation

The cloud infrastructure should automatically minimize the costs of:

- deploying, operating and evolving the *ad hoc* cloud so that it is both highly-available to *ad hoc* cloud users and non-disruptive to primary users
- ensuring that the execution of applications is efficient and reliable, with high probability, over long periods

This requires the *ad hoc* cloud to exhibit autonomic capabilities [14]. It should automatically adapt the extent of each cloudlet and the placement of its data and computation, driven by appropriate management policies. The requirement poses a significant scientific challenge, since it involves an understanding of the cost/benefit ratios associated with any given policy on resource allocation. Dynamic, multi-objective optimisation is necessary in order to coordinate resource utilisation policies, and to evolve policies in response to changing circumstances—about which knowledge is typically scarce.

Because an *ad hoc* cloud runs on non-dedicated resources, harvesting and harnessing activities must remain acceptably non-intrusive, requiring stringent constraints on policies and their actions. Moreover, since the usage patterns of machine owners may vary widely, policies must be informed by efficient, scalable monitoring and performance modelling from which reliable, robust cost/benefit ratios can be derived.

## 4.3 Quality Of Service

To deliver QoS guarantees, two separate services are required: to make policy decisions based on QoS negotiations with external parties; and to provide mechanism to implement policy decisions. We propose to use the *Broker* and *Dispatcher* patterns [16]. These are embodied as distributed services that are structured in the same way as user-level services, making them autonomic and capable of changing their behaviour and resource usage in response to changing request patterns.

A **broker** establishes and manages up-front agreements with users of the cloud, for the provision of services at certain QoS levels for certain periods. It matches reservation requests against expected available resources in tandem with other commitments, and informs the requester whether or not the request can be satisfied. Where the broker reaches an agreement, it seeks to ensure that resources are pre-configured (e.g. with suitable service deployments) in a way that enables the agreement to be met.

The requirements of different types of service may need to be coordinated to meet QoS goals. The broker identifies whether the resource requirements and their associated constraints can be met. Such assessment requires access to information on the available computational resources, their historic loads and availabilities and other commitments that relate to them. A search must be made for a future configuration that is predicted to meet the requirements of the current request and future requests.

A **dispatcher** exposes the user-level cloud services. Where a request is made to access a service for which there is an established agreement, the dispatcher makes use of the resources reserved by the broker in anticipation of the request. Where there is no pre-arranged agreement, the request is still directed to a relevant service by the dispatcher on a best-effort basis.

# 5. EXAMPLE

As an example of this architecture we describe the H2O database system [15]. H2O is a relational database based on the open source H2 system [19] that is intended for deployment on an *ad hoc* cloud. It offers a full set of relational operations including a user interface and JDBC linkage. Unlike traditional desktop relational systems such as MySQL, the system resides within heterogeneous desktop systems hosted within an enterprise. To understand how the H2O system maps onto the architecture described in this paper, two components of H2O must be considered separately: those corresponding to the cloudlet, and those corresponding to the cloud elements.

The cloud elements each run a complete relational database engine which is responsible for whole or partial relational tables from the database. This corresponds to the engine component shown in Figure 2. A Java abstract machine, augmented by a restricted interface to local persistent storage, hosts each database engine. The modeller/manager element tracks the amount of persistent storage used on the node and the bandwidth usage etc. The programs that are executed by the engine are SQL fragments that are delivered to the node either from a local user interface or sent by another cloud element. In addition to the network interface exporting SQL functionality, a Java RMI interface provides functionality for configuring both individual cloud elements and deployments on each node.

The cloudlet component is required to track the individual elements of the database and maintain the database metadata. For example, individual relations are autonomically replicated across multiple cloud elements for resilience. Consequently, when a query attempts to access a relation for the first time or following a failure, it must query the cloudlet to bind to the managers running within an individual cloud element. This functionality is achieved by running a distributed database manager in the cloudlet above a P2P infrastructure, with each of the individual components of the cloudlet being hosted by cloud elements.

The cloud infrastructure on each node provides the ability to instantiate cloud elements on individual nodes. We have developed technology in earlier systems [7] that permits securely signed bundles of code and data to be instantiated on machines.

# 6. CONCLUSIONS

The *ad hoc* cloud model could allow complex cloud-style applications to exploit untapped resources on non-dedicated hardware. We believe that this approach has the potential to:

- enable organizations to reduce IT costs;
- enable organizations to obtain the benefits of cloud computing in new application areas;
- reduce net energy consumption by IT activities.

We have outlined a case for ad hoc cloud computing, a set of resulting research challenges, and as a starting point, a proposed architecture.